\documentclass[aps,pre,twocolumn,superscriptaddress,showpacs]{revtex4}
\usepackage{amssymb}
\usepackage{epsfig}
\usepackage{amsmath}
\usepackage{times}

\begin{document}

\title{Evolutionary Subnetworks in Complex Systems}

\author{Menghui Li}
\affiliation{Temasek Laboratories, National University of
Singapore, 117508, Singapore} \affiliation{Beijing-Hong
Kong-Singapore Joint Centre for Nonlinear \& Complex Systems
(Singapore), National University of Singapore, Kent Ridge, 119260,
Singapore}
\author{Xingang Wang}
\email[Corresponding author. Email address: ]{wangxg@zju.edu.cn}
\affiliation{Institute for Fusion Theory and Simulation, Zhejiang
University, Hangzhou, China 310027}
\author{Choy-Heng Lai}
\affiliation{Beijing-Hong Kong-Singapore Joint Centre for
Nonlinear \& Complex Systems (Singapore), National University of
Singapore, Kent Ridge, 119260, Singapore} \affiliation{Department
of Physics, National University of Singapore, Singapore 117542}

\begin{abstract}
Links in a practical network may have different functions, which
makes the original network a combination of some functional
subnetworks. Here, by a model of coupled oscillators, we investigate
how such functional subnetworks are evolved and developed according
to the network structure and dynamics. In particular, we study the
case of evolutionary clustered networks in which the function of
each link (either attractive or repulsive coupling) is updated by
the local dynamics. It is found that, during the process of system
evolution, the network is gradually stabilized into a particular
form in which the attractive (repulsive) subnetwork consists only
the intralinks (interlinks). Based on the properties of subnetwork
evolution, we also propose a new algorithm for network partition
which is distinguished by the convenient operation and fast
computing speed.

\end{abstract}
\date{\today }
 \maketitle

The past decade has witnessed the blooming of network science, in
which one important issue is to explore the interplay between the
network structure and dynamics \cite{CN:REV,SYN:REV}. While the
influences of the network structure on dynamics have been
intensively studied in the past \cite{SYN:NET}, recently attentions
have also been paid to the influences of the network dynamics on
structure, i.e. the evolution of complex networks driven by dynamics
\cite{BBH:2004,GL:2004,ADP:2006,ZK:2006,BILPR:2007,OCKK:2008,LANBHB:2008}.
In Ref. \cite{GL:2004} it has been shown that, rewiring network
links according to the node synchronization, a random network can be
gradually developed to a small-world network. In Ref. \cite{ZK:2006}
it has been shown that, driven by node synchronization, the weight
of the network links can be developed to a particular form in favor
of global network synchronization. Besides network evolution,
dynamics has been also used for network detection, e.g., detecting
the modular structures in clustered complex networks
\cite{ADP:2006,ZK:2006,BILPR:2007,OCKK:2008,LANBHB:2008,ZZZHK:2006}.

It has been well recognized that links in a practical network are
usually different from each other. In previous studies, this has
been mainly reflected in the variation of the weight of the network
links, i.e., the weighted network \cite{CN:REV}. Weighted network,
however, describes only the case of single-function networks, i.e.
all links in the network have the same function, but failing to
describe the situation of multi-function networks in which the
network links have the diverse functions. A type of commonly seen
multi-function networks in practice is the cooperation-competition
network (CCN) \cite{NERVOUS:BOOK,PDG}, in which the network links
are divided into two groups of opposite functions. For instance, in
the nervous network of the human brain, the synapses are roughly
divided into two groups, excitatory and inhibitory, which play the
contrary roles to the neuron activities \cite{NERVOUS:BOOK}. Another
typical example of CCN is the relationship network shown in the
prisoner's dilemma game, in which each suspect may either cooperate
with (remain silent) or defect from (betray) the other suspects
\cite{PDG}.

For multi-function networks like CCN, to facilitate the analysis, it
will be convenient if we treat the different groups of links
separately. That is, we pick out links serving the same function
and, together with their associated nodes, construct a small
single-function network. In this way, a multi-function network can
be decomposed into a number of functional subnetworks, while each
supports a unique function to the system behaviors. Here an
interesting question is: How do these functional subnetworks
co-evolve with each other and develope into their ``adult" forms
according to the system properties, e.g., the network structure and
dynamics?

To mimic the evolution of the functional subnetworks, we propose the
following model of coupled phase oscillators,
\begin{equation}
\dot{\theta}_i = \omega_i + \varepsilon \sum_{j=1}^{N}a_{ij} [ \sin
(\theta_j - \theta_i)e_{ij} + \cos(\theta_j - \theta_i)(1-e_{ij}) ].
\label{model}
\end{equation}
Here, $i,j=1,2,\ldots,N$ are the node indices, $\varepsilon$ is the
uniform coupling strength. $\theta_i$ and $\omega_i$ are the instant
phase and intrinsic frequency of the $i$th oscillator, respectively.
The network structure is represented by the adjacency matrix $A=\{
a_{ij} \}$, in which $a_{ij}=1$ if nodes $i$ and $j$ are directly
connected, and $a_{ij}=0$ otherwise. $E(t)=\{ e_{ij}(t) \}$ is a
time-dependent binary matrix whose elements are defined as follows.
Let $\psi_i(t)$ be the instant phase of the local order parameter
defined by the equation \cite{ROH:2005}
\begin{equation}
r_i e^{i\psi_i(t)}=\sum_{j=1}^N a_{ij}e^{i\theta_j(t)}.
\end{equation}
We set $e_{ij}(t)=1$ if the difference between $\psi_i$ and $\psi_j$
is smaller than a threshold $D$, otherwise we set $e_{ij}(t)=0$.

Different from the traditional models of coupled phase oscillators,
in Eq. (\ref{model}) the coupling term is made up of two parts of
the opposite functions. While the attractive coupling,
$H_A=\sin(\theta_j-\theta_i)$, is going to synchronize the connected
nodes, the repulsive coupling, $H_R=\cos(\theta_j-\theta_i)$, will
work against this tendency. These opposite functions, however,
cannot coexist. That is, at any time instant each link can only take
on one type of coupling function, either attractive ($e_{ij}=1$) or
repulsive ($e_{ij}=0$). The attractive links, together with their
associated nodes, constitute the attractive subnetwork, which is
represented by the matrix $B=A\circ E$ ($``\circ"$ is the entry-wise
product and $I$ is the identity matrix). Similarly, we can construct
the repulsive subnetwork, and represent it by the matrix $R=A\circ
(I-E)$. Because $e_{ij}(t)$ is being updated with the system
dynamics, the two subnetworks, therefore, will also be changing with
time. It should be noted that, despite the evolution of the
subnetworks, the global network structure is kept unchanged, i.e.,
$B(t)+R(t)\equiv A$. Imagine a complex network that is weakly
coupled and there is no synchronization between any pair of nodes.
It can be expected that, as the system evolves, the two subnetworks
will be continuously updated in a random fashion. The question we
are interested here is: What happens to the evolution of the
subnetworks if the coupling strength is stronger?

We start our investigation by considering the evolution of clustered
networks (CN) \cite{CN:REV}. A typical model of CN is the \emph{ad
hoc} network introduced in Ref. \cite{FOOTBALL}, which consists of
$4$ clusters, each contains $32$ nodes. In this model, each node on
average has $\left< k \right>=16$ links, among which $\left< k_l
\right>$ links are connected to nodes within the same cluster, i.e.
the intralinks, and $\left< k_p \right>$ links are connected to
nodes from different clusters, i.e, the interlinks. Since we are
interested in the case of strongly coupled clustered networks, we
use, without loss of generality, in our simulations the parameters
$\varepsilon =5$ and $\left< k_p \right>=1$. Meanwhile, to generate
the matrix $E(t)$, we use the threshold $D=0.45$. (The influences of
these parameters to the evolution will be discussed later.) The
natural frequencies and initial conditions of the oscillators are
chosen randomly from the ranges $[0,1]$ and $[0, 2\pi)$,
respectively. To monitor the evolution, we keep a record of the
instant states of the oscillators ($\{\theta_i(t)\}$) and the
instant subnetwork matrices ($B(t)$ and $(R(t)$).

\begin{figure}[tbp]
\begin{center}
\includegraphics[width=\linewidth]{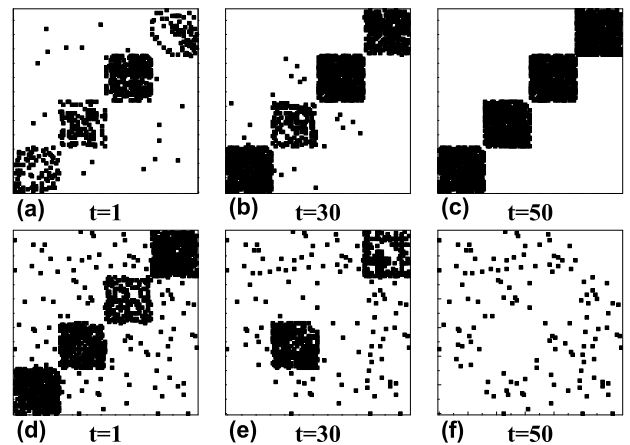}
\caption{The evolution of functional subnetworks in the \emph{ad
hoc} network. (a)-(c) The evolution of the attractive matrix $B(t)$.
(d)-(f) The evolution of the repulsive matrix $R(t)$. From time
$t\approx50$, all intralinks (interlinks) are contained in the
attractive (repulsive) subnetwork, and the structures of the
subnetworks will be stabilized. Nodes are rearranged according to
the topological clusters.} \label{Fig_1}
\end{center}
\end{figure}

The evolution of the structures of the subnetworks can be described
as follows. At the beginning, the two subnetworks have similar
configuration, i.e. both are an abbreviated version of the original
network [Figs. 1(a) and (d)]. This is because in a very short time
the oscillators have not reached any synchronization, and therefore
the matrices $B$ and $R$ are mainly determined by the initial
conditions of the oscillators. But, due to the small value of $D$,
the repulsive subnetwork has more links than the attractive
subnetwork. Then, as time increases, the \emph{interlinks} are
gradually excluded from the attractive subnetwork; meanwhile, the
\emph{intralinks} are excluded from the repulsive subnetwork [Figs.
1(b) and (e)]. The separation of the two subnetworks, however, is
not an even process, as some links may jump between the subnetworks
repeatedly before settling down. Finally, at the time about $T= 50$,
the subnetworks are stabilized into fixed structures and the
evolution is complete. In this final stationary state, all
intralinks (intralinks) of the network are included in the
attractive (repulsive) subnetwork [Figs. 1(c) and (f)].

\begin{figure}[tbp]
\begin{center}
\includegraphics[width=\linewidth]{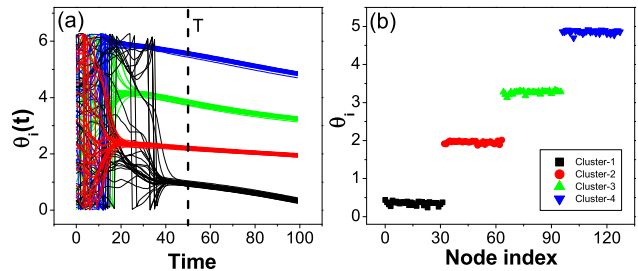}
\caption{(Color online) For the same network as in Fig. 1, (a) the
time evolution of the oscillator states and (b) a snapshot of the
oscillator states at time $t=100$.} \label{Fig_2}
\end{center}
\end{figure}

With the evolution of the subnetworks, the system dynamics is also
changed. As shown in Fig. 2(a), with the increase of time, the
oscillators are gradually organized into $4$ synchronous clusters.
The pattern of the synchronous clusters is more evident in Fig.
2(b), where a snapshot of the oscillator states is taken at time
$t=100$. After a check of the node indices of the pattern, it is
found that the organization of the dynamical clusters obey precisely
the topological clusters. Moreover, in forming the synchronous
clusters, it is found that the overlapping nodes, i,e. nodes which
have interlinks, are more difficult to be synchronized than those
internal nodes. This is clearly evidenced in the synchronizing
process of the first cluster, where the few overlapping nodes are
traveling among the synchronous clusters for an extra period before
settling down (the black curves in Fig. 1(a)). Another interesting
finding is that, in the stabilized pattern, the states of the
synchronous clusters are well separated from each other, which, of
course, is attributed to the repulsive coupling on the interlinks.

A prominent feature of our evolutionary model is that the link
functions are updated only by the local network information, i.e.,
the phase of the local order parameter. While it has been commonly
believed that the identification of the link attribute, i.e.
intralink or interlink, relies on the knowledge of the global
network information, e.g., the betweenness centrality of the network
links, it is somewhat surprising to see that here the identification
can be accomplished by only the local network information. This
interesting phenomenon can be explained by a local mean-field
theory, as follows. Let $a_{ij}$ be an intralink of a clustered
network. Since nodes inside a cluster are densely connected, nodes
$i$ and $j$ thus are surrounded by a similar set of neighboring
nodes. Because of the large overlap of their neighboring sets, the
average phases $\psi_i$ and $\psi_j$ will have small difference,
leading to the attractive coupling on the intralink, i.e.,
$e_{ij}=1$. In contrast, if $a_{ij}$ is an interlink, the nodes $i$
and $j$ will be surrounded by very different neighboring sets, which
will generate a larger difference in the average phase, finally
leading to the repulsive coupling on the interlink.

We next discuss the influences of the network structure on the
evolution. Having understood the critical role of the overlapping
neighbors in the evolution, we are able to predict that the above
phenomena of subnetwork formation can be observed in any network of
clear modular structures. To verify this, we have studied the
evolution of the other two typical network models. The first one is
the overlapping clustered network studied in Ref.
\cite{LANBHB:2008}, in which two larger clusters (each has $96$
nodes) are mediated by two smaller complete clusters (each has $4$
nodes). The smaller clusters have no direct connection, but each is
connected to the two larger clusters by an equal number of links. By
analyzing their neighbor sets, the network nodes are immediately
classified into four groups: two for the larger clusters and two for
the smaller clusters. Correspondingly, the oscillators are expected
to be synchronized into $4$ dynamical clusters. This is indeed what
we have observed in the simulations [Fig. 3(a) and (b)]. The second
model we have simulated is an ER network \cite{CN:REV}. Since an ER
network has no topological cluster, the neighboring sets of the
network nodes thus are different from each other. According to the
neighbor-set analysis, this will lead to the repulsive couplings on
the links, and generating the turbulent system dynamics. This is
indeed what we have found at the beginning of the evolution [Fig.
3(c) and (d)]. The repulsive network and turbulent dynamics,
however, are unstable. As shown in Fig. 3, after a transient period,
the repulsive couplings are quickly switched to the attractive
couplings and the turbulent state is changed to the state of global
synchronization [Figs. 3(c) and (d)]. The switching of the link
functions and system dynamics suggest the dual properties of the ER
network, i.e., it can be regarded either as containing no module or
as containing one unified module.

\begin{figure}[tbp]
\begin{center}
\includegraphics[width=\linewidth]{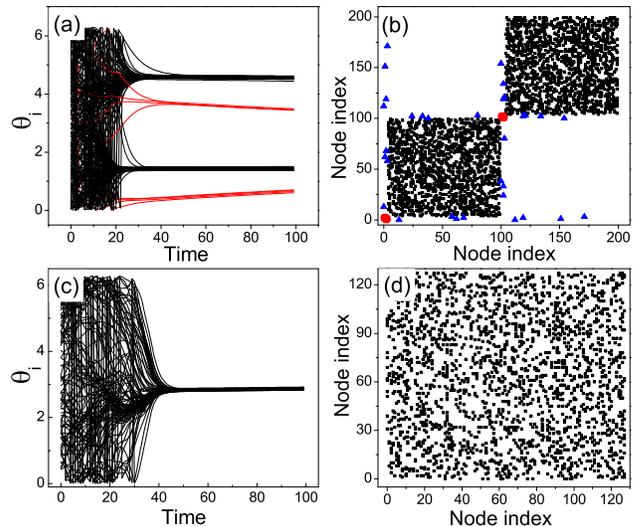}
\caption{(Color online) For the overlapped clustered network, (a)
the evolution of the system dynamics and (b) the stabilized link
functions. In (b), the intralinks of the larger and smaller clusters
are marked by black squares and red dots, respectively. The
interlinks are marked by blue triangles. For a ER network of size
$N=128$ and average degree $\left< k \right >=16$, (c) the evolution
of the system dynamics and (d) the stabilized link functions. The
switching of the link functions and system dynamics are started at
about $t=30$. In (d), all network links are granted with the
attractive coupling function.} \label{Fig_3}
\end{center}
\end{figure}

We go on to study the influences of other system parameters on the
evolution, including the coupling strength $\varepsilon$, the
threshold $D$, and the local dynamics. The numerical results show
that, given the network has a clear modular structure, the link
attribute can always be successfully detected by the local dynamics,
despite the changes of $\varepsilon$ and $D$. Specifically, for the
\emph{ad hoc} network of Fig. 1, the system will always develop to
the same functional subnetworks [Fig. 1] and dynamical pattern [Fig.
2] in the parameter space constructed by $\varepsilon \in [0.5,8]$
and $D \in [0.45,\pi/2]$. Furthermore, the main feature of the
evolution is independent of the specific form of the local dynamics,
as has been verified by other nonlinear oscillators, such as the
Logistic map and Lorenz oscillator. However, it should be pointed
out that, by changing these parameters, the transient process of the
network evolution could be strongly affected, e.g., the transient
states of the evolution.

Finally, we discuss the possible application of the evolutionary
model in network partition
\cite{FOOTBALL,GDGGA:2003,RB:2004,CNM:2004,WL:2008}. In network
partition, the performance of an algorithm is mainly measured by the
following three factors: ease of implementation, accurate detection
and fast computation. All these factors are well met in the
evolutionary subnetwork model (ESM). Firstly, on the aspect of ease
of implementation, the model employs only the local network
information, and the partition is accomplished automatically by the
system dynamics. In particular, at the end of the evolution, based
on the states of the synchronous clusters, the topological clusters
can be readily identified. Automatic detection and local network
information are the main features of the ESM algorithm, which are
also the major difference to the other dynamics-based algorithms
\cite{BILPR:2007,OCKK:2008,RB:2004}. Secondly, the ESM algorithm is
fast in computing speed. The computational cost of the ESM method is
estimated to be $O(NT)$, with $N$ the network size and $T$ the
transient time of the evolution. To estimate the computational cost
further, it is necessary to characterize the relationship between
$T$ and $N$. Numerically, we have checked this relationship by
increasing the size of the \emph{ad hoc} network in the following
two approaches. In the first approach, the number of the clusters
are kept unchanged, but the size of the $4$th cluster is gradually
increased from $32$ to $864$. In the second approach, the size of
each cluster is kept unchanged, but the number of the clusters is
increased from $4$ to $30$. The variations of $T$ as a function of
$N$ are plotted in Fig. 4(a), together with the synchronizing time,
$T'$ , of the largest cluster in the network. Very interestingly, it
is found that for both approaches we have $T\propto T'$. That is,
the transient time of the whole network is proportional to that of
the largest cluster. Particularly, for the first approach we even
have $T\sim T'$. Previous studies have indicated that $T'$ is mainly
determined by the intrinsic properties of the cluster, e.g. the
cluster size, instead of the global network properties
\cite{WHLL:2007}. This implies that $T'\sim M_{max}$, with $M_{max}$
the size of the largest cluster in the network. Therefore, the
computational cost of ESM is estimated to be proportional to
$O(NM_{max})$. Since for practical networks we generally have
$M_{max} \ll N$, the computational cost of ESM thus is estimated to
increase linearly with the network size. Finally, on the aspect of
detecting accuracy, the ESM algorithm works very well for clustered
networks and reasonably well for fuzzy networks [Fig. 4(b)].

\begin{figure}[tbp]
\begin{center}
\includegraphics[width=\linewidth]{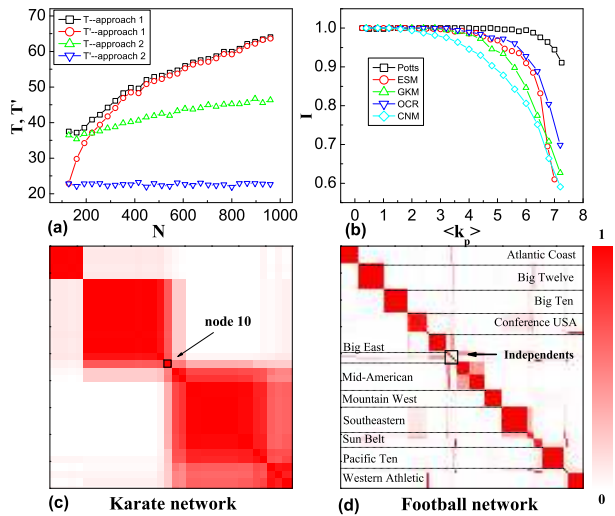}
\caption{(Color online) For the \emph{ad hoc} network of Fig. 1, (a)
the transient time of the whole network, $T$, and the synchronizing
time of the largest cluster in the nework, $T'$, as a function of
the network size, and (b) the variation of the mutual information
(see Ref. \cite{DGDA:2005} for details about mutual information) as
a function of $\left<k_p\right>$ generated by the partition
algorithms ``Potts" \cite{RB:2004}, ``ESM", ``GKM" \cite{OCKK:2008},
``OCR" \cite{BILPR:2007}, and ``CNM" \cite{CNM:2004}. The
coappearance matrix generated by ESM for (c) the Zachary's karate
club and (d) the football club. The overlapping nodes are marked by
the squares. In (c) and (d), each data is averaged over $100$
realizations. } \label{Fig_4}
\end{center}
\end{figure}

As applications of the new algorithm, we have tested the partitions
of two empirical clustered networks. The first one is the Zachary's
karate club network \cite{KARATE}, which contains $34$ nodes and
$78$ links. The numerical result is plotted in Fig. 4(c), in which
the network is clearly divided into $4$ clusters. Moreover, the
overlapping node, i.e. the $10$th node in the network, is also well
characterized. These results coincide with that of Ref.
\cite{KARATE}. The second example we have tested is the football
network \cite{FOOTBALL}, which contains $115$ nodes (teams) and
$613$ (matches) links. The numerical result is plotted in Fig. 4(d),
in which the network is clearly divided into a number of conferences
(clusters). Again, the independent teams, which have equal number of
games (links) with multiple conferences, are well identified by the
overlapping nodes. These results coincide with that of Ref.
\cite{FOOTBALL}.

In summary, we have studied the evolution of functional subnetworks
in clustered networks and proposed a new algorithm for partitioning
networks. Hopefully, the finding that the network function is
jointly determined by the network structure and system dynamics
could be helpful to the study of complex behaviors in functional
networks.



\begin{thebibliography}{99}

\bibitem{CN:REV} R. Albert and A.-L. Barab\'{a}si, Rev. Mod. Phys.
\textbf{74}, 47 (2002); M.E.J. Newman, SIAM Rev. \textbf{45}, 167
(2003).

\bibitem{SYN:REV} S. Boccaletti, \emph{et.al.}, Phys. Rep. \textbf{424}, 175 (2006); A. Arenas, \emph{et.al.}, Phys. Rep.
\textbf{469}, 93 (2008).

\bibitem{SYN:NET} X. F. Wang and G. Chen, Int. J. Bifurcation Chaos Appl.
Sci. Eng. \textbf{12}, 187 (2002); M. Barahona and L. M. Pecora,
Phys. Rev. Lett. \textbf{89}, 054101 (2002); T. Nishikawa,
\emph{et.al.}, Phys. Rev. Lett. \textbf{91}, 014101 (2003); X. G.
Wang, \emph{et.al.}, Phys. Rev. E \textbf{75}, 056205 (2007); A.
Arenas, Phys. Rev. Lett. \textbf{98}, 034101 (2007).
%


\bibitem{BBH:2004} I.V. Belykh, \emph{et.al.}, Physica D \textbf{195}, 188
(2004); W. Li, \emph{et.al.}, Phys. Rev. E \textbf{76}, 045102(R)
(2007).

\bibitem{GL:2004} P. Gong, C.V. Leeuwen, Europhys. Lett.
\textbf{67}, 328 (2004).

\bibitem{ADP:2006} A. Arenas, \emph{et.al.},
Phys. Rev. Lett. \textbf{96}, 114102, (2006)

\bibitem{ZK:2006} C. Zhou and J. Kurths, Phys. Rev. Lett.
\textbf{96}, 164102 (2006).

\bibitem{BILPR:2007} S. Boccaletti, \emph{et.al.}, Phys. Rev. E \textbf{75}, 045102(R)
(2007).

\bibitem{OCKK:2008} E. Oh, \emph{et.al.}, Europhys.
Lett. \textbf{83}, 68003 (2008).

\bibitem{LANBHB:2008} D. Li, \emph{et.al.},
Phys. Rev. Lett. \textbf{101}, 168701 (2008).


\bibitem{ZZZHK:2006} C.S. Zhou, \emph{et.al.}, Phys. Rev. Lett. \textbf{97}, 238103
(2006).

\bibitem{NERVOUS:BOOK} D. Purves and J. Lichtman, \emph{Principles of Neural Development} (Sinauer
Associates, Sunderland, MA, 1985).

\bibitem{PDG}J. Hofbauer and K. Sigmund, \emph{Evolutionary Games and Population
Dynamics} (Cambridge University Press, Cambridge, 1998).

\bibitem{ROH:2005}J. G. Restrepo, \emph{et.al.}, Phys. Rev. E \textbf{71}, 036151
(2005); X.G. Wang, \emph{et.al.}, Chaos \textbf{18}, 037117 (2008).

\bibitem{FOOTBALL} M. Girvan and M.E.J. Newman, Proc. Natl. Acad.
Sci. USA \textbf{99}, 7821 (2002).

\bibitem{GDGGA:2003}R. Guimer\'{a}, \emph{et.al.}, Phys. Rev. E \textbf{68}, 065103(R), (2003); J. Duch and A.
Arenas, Phys. Rev. E \textbf{72}, 027104, (2005);

\bibitem{RB:2004} J. Reichardt and S. Bornholdt, Phys. Rev. Lett.
\textbf{93}, 218701 (2004).

\bibitem{CNM:2004} A. Clauset, \emph{et.al.}, Phys.
Rev. E \textbf{70}, 066111 (2004).


\bibitem{WL:2008} J. Wang, C.-H. Lai, New J. Phys. \textbf{10},
123023 (2008).

\bibitem{DGDA:2005} L. Danon, \emph{et.al.}, J. Stat. Mech. P09008 (2005).

\bibitem{WHLL:2007} X.G. Wang, \emph{et.al.}, Phys. Rev. E
\textbf{76}, 056113 (2007).

\bibitem{KARATE} W.W. Zachary, J. Anthropol. Res. \textbf{33},452
(1977).



\end{thebibliography}
\end{document}